%% file: main.tex
\documentclass[runningheads]{llncs}

\usepackage{graphicx}
\usepackage{booktabs}
\usepackage{amsmath}
\usepackage{siunitx}
\usepackage[table]{xcolor}
\usepackage{tcolorbox}
\usepackage{listings}
\lstdefinestyle{codeassay}{%
  basicstyle=\ttfamily\footnotesize,
  keywordstyle=\bfseries,
  commentstyle=\itshape\color{gray},
  showstringspaces=false,
  breaklines=true,
  columns=fullflexible,
  frame=single,
  rulecolor=\color{gray},
  framesep=4pt,
  language=Python,
  aboveskip=2pt, belowskip=2pt,
}
\usepackage{url}
\usepackage{hyperref}

\newcommand{\passk}{pass@$k$}

\hypersetup{
    colorlinks=true,
    linkcolor=blue,   
    citecolor=blue,   
    urlcolor=blue     
}

\usepackage{orcidlink}
\usepackage{tabularx}

\usepackage{fancyhdr}

\fancypagestyle{firstpage}{
    \fancyhf{}
    \fancyhead[L]{\fontsize{6.5pt}{9pt}\selectfont \textit{PROFES 2026: 27th International Conference on Product-Focused Software Process Improvement}}
    
}

\begin{document}

\title{CodeAssay: A Multi-Metric Benchmark with Audited Ground Truth for LLM Code Generation}

\titlerunning{CodeAssay: A Multi-Metric, Audited Code-Generation Benchmark}

\author{Shahbaz Siddeeq\inst{*,1}\orcidlink{0009-0003-9030-8841} \and
Muhammad Waseem\inst{1}\orcidlink{0000-0001-7488-2577} \and
Umar Subhan Malhi\inst{2}\orcidlink{0000-0002-1804-3573} \and
Pekka Abrahamsson\inst{1}\orcidlink{0000-0002-4360-2226}}
\authorrunning{S. Siddeeq et al.}

\institute{Tampere University, Tampere, Finland\\
\email{\{shahbaz.siddeeq,muhammad.waseem,pekka.abrahamsson\}@tuni.fi} \and
Stony Brook University, Stony Brook, NY 11794, United States of America \\
\email{umar.malhi@stonybrook.edu}}

\maketitle

\thispagestyle{firstpage}
\pagestyle{headings}

\begingroup
\renewcommand{\thefootnote}{*}
\footnotetext{Corresponding author: Shahbaz Siddeeq \textless\textit{shahbaz.siddeeq@tuni.fi}\textgreater}
\endgroup

\begin{abstract} Large Language Models are increasingly evaluated for code generation using test-based benchmarks. The validity of such evaluations depends on the reliability of their references and tests, while test-based correctness captures only part of the observable properties of generated code. We present CodeAssay, a taxonomy-first benchmark of 185 Python tasks across ten software-engineering categories. It combines audited ground truth, public tests for generation and repair, hidden tests for grading, mutation-based test-suite validation, and selected code-property measures. Regrading fixed model outputs after the audit changed 170 of 1,890 correctness labels (9.0\%) and increased the measured best-to-worst model spread from 11.9 to 23.7 percentage points, although aggregate correctness remained nearly unchanged. The complete and hidden test suites achieved mutation scores of 82.6\% and 74.8\%, respectively. Across seven proprietary LLMs, standard-prompt correctness ranged from 77.3\% to 98.9\%, with significant differences in 12 of 21 model pairs. On the 120 tasks solved by all 14 model-prompt configurations, no model performed best across all selected code properties. A security-focused prompt produced no significant change in correctness or consistent reduction in the selected static-analysis findings, while increasing program length and cyclomatic complexity across all models. These findings show that reliable evaluation of LLM-generated code requires validated ground truth, protected tests, and multiple explicitly interpreted measures. CodeAssay provides a reproducible basis for evidence-based model evaluation in AI-augmented software development. \end{abstract}

\keywords{Code generation \and Large language models \and Software-quality
benchmark \and Mutation testing \and Empirical software engineering.}

\input{Introduction}
\input{Background}
\input{CodeAssay}
\input{ExperimentalSetup}
\input{Results}
\input{Discussion}
\input{ThreatsValidity}
\input{Conclusion}

\bibliographystyle{splncs04}
\bibliography{references}

\end{document}

%% file: Introduction.tex
\section{Introduction}
\label{sec:intro}

Benchmarks are used to evaluate and compare Large Language Models (LLMs) on code generation~\cite{hou2024llm4se}. A comparison drawn from one is only as reliable as the properties it measures and the reference solutions and test suites it grades against. Two concerns follow. Do current benchmarks assess relevant properties of generated programs, and is the ground truth they grade against reliable?

Established code-generation benchmarks evaluate generated programs primarily by whether they pass a test suite. HumanEval scores functional correctness through the \passk{} estimator over 164 problems~\cite{humaneval2021}, MBPP pairs each prompt with a reference solution and three test cases~\cite{mbpp2021}, BigCodeBench extends the setting to library-rich tasks across seven domains~\cite{bigcodebench2024}, and CoderEval targets non-standalone functions drawn from real projects~\cite{codereval}. Functional correctness is necessary, but it does not characterize every observable property of a generated program. Programs that pass the same tests can differ in length, cyclomatic complexity, style violations per logical line, and the constructions flagged by a static analyzer. RACE evaluates readability, maintainability, and efficiency alongside correctness~\cite{race2024}. These properties are measurable and worth reporting, but none is a complete measure of maintainability, security, or software product quality.
The second concern is the reliability of benchmark ground  truth. EvalPlus reports incorrect ground truth in HumanEval, finding 18 defects in \SI{11}{\percent} of problems, and shows that the original test inputs failed to expose plausible-but-wrong programs~\cite{evalplus2023}. Where a reference is wrong or a suite is permissive, scores computed against it may be distorted, and the distortion is invisible to anyone who adopts the benchmark as given. This concerns benchmark validity rather than any particular model.

We present CodeAssay, a Python code-generation benchmark that addresses ground-truth reliability through its construction and validation while broadening evaluation beyond functional correctness. It is intended to support model selection and quality-gate design for teams adopting LLMs, with reproducible measurements and grading. We make the following contributions:

\begin{itemize}

    \item We develop and validate CodeAssay, a taxonomy-first Python code-generation benchmark comprising 185 tasks across ten categories  informed by SWEBOK~v4~\cite{swebok4}. It combines independently audited ground truth, public tests for generation and repair, hidden tests for grading, mutation-based test-suite validation, and fixed-program regrading analysis (Sections~\ref{sec:construction} -~\ref{sec:validation}).

    \item We evaluate CodeAssay using seven proprietary LLMs under standard and security-focused prompts. We compare their functional correctness and, on the tasks solved correctly by all 14 model--prompt configurations, selected code properties (Section~\ref{sec:results}).

    \item We open-source CodeAssay and its replication package, including the tasks, tests, prompts, generated programs, validation artifacts, and analysis scripts~\cite{codeassay2026artifact}.

\end{itemize}

%% file: Background.tex
\section{Related Work}
\label{sec:related}

Several benchmarks provide particularly relevant methodological
comparisons with CodeAssay.
ClassEval~\cite{classeval} works at class level and reports statement and branch
coverage above \SI{98}{\percent}, together with a mutation score of
\SI{83.7}{\percent}, measured over its full suites rather than over the subset
that determines a grade. 
LiveCodeBench~\cite{livecodebench2024} mitigates contamination
by continually incorporating newly released contest problems and filtering them
according to model cutoff dates.
EvalPlus~\cite{evalplus2023} strengthens an existing suite rather than authoring
new tasks. It generates seed inputs with an LLM, expands them through type-aware
mutation, and optionally reduces the resulting suites while
preserving branch coverage, mutant kills, and empirical LLM-sample kills.
RACE~\cite{race2024} adapts tasks from HumanEval+, MBPP+,
ClassEval, and LeetCode and adds quality-specific requirements, rather than
authoring a new taxonomy-first task set.
\begin{figure}[!b]
  \centering
  \includegraphics[width=\linewidth, clip, trim=20 10 20 10]{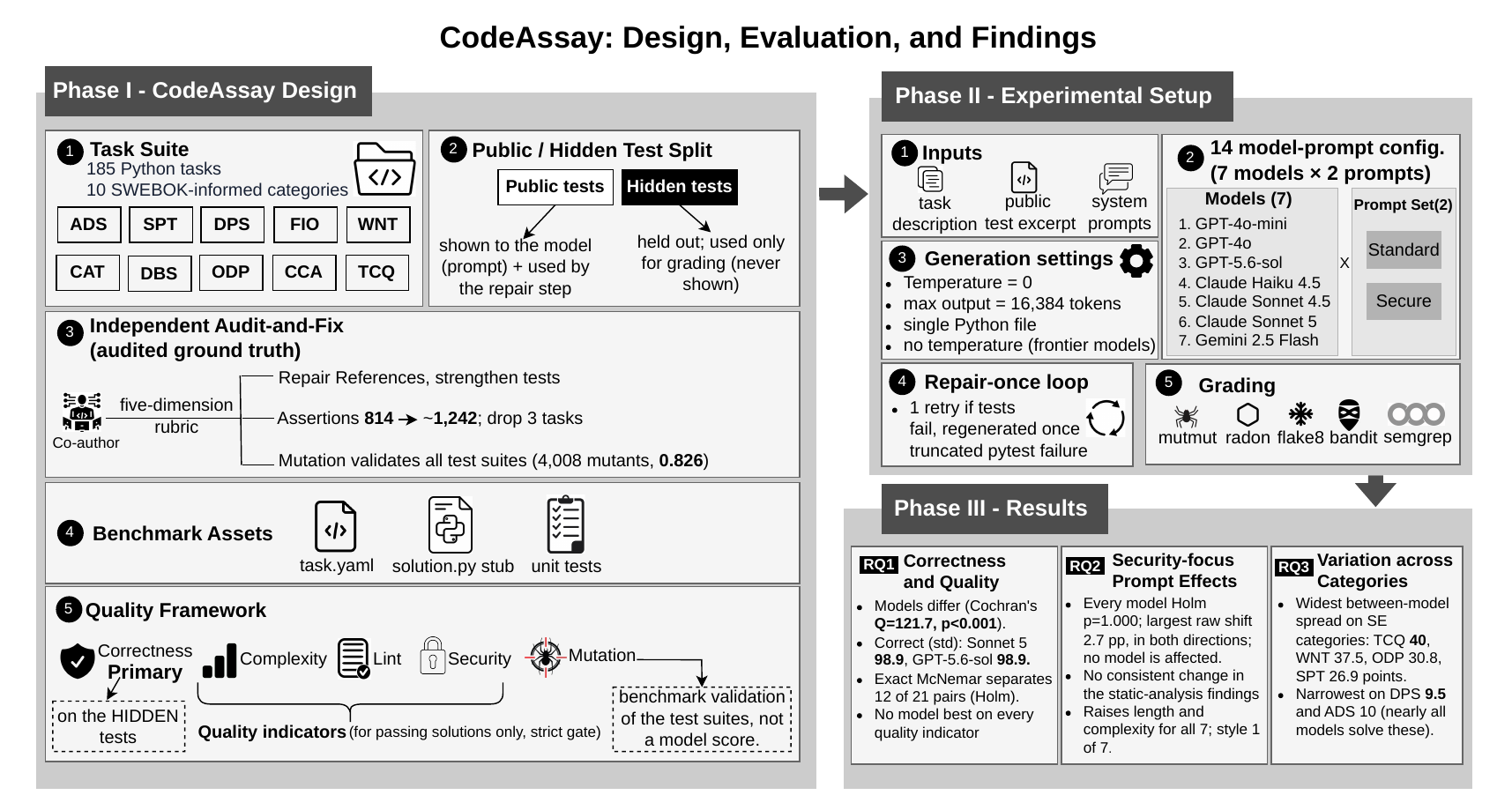}
  \caption{CodeAssay in three phases: benchmark design and audit, experimental
  setup, and results.}
  \label{fig:pipeline}
\end{figure}

A review of 109 studies finds that quality characteristics of generated code
beyond functional correctness are under-studied~\cite{sun2026}. Security is
benchmarked separately from correctness by prompt datasets such as
SecurityEval~\cite{siddiq2022securityeval} and by
BaxBench~\cite{vero2025baxbench}, which reports exploits succeeding against
programs that pass their tests.
CodeAssay includes no task-specific security oracle and reports
only selected findings from Bandit and Semgrep.
Pearce et al.~\cite{pearce2022} found that code written with an AI assistant
often contained known weakness patterns, and Perry et
al.~\cite{perry2023insecure} observed that developers using such an assistant
wrote less secure code while believing that code more secure. We therefore
report selected static-analysis findings separately rather than folding them
into functional correctness.

Auditing benchmark validity has precedent beyond EvalPlus. A
re-examination of SWE-bench~\cite{xue2026swebenchplus} identified solution
leakage and tests too weak to verify some patches; reclassifying the affected
successful patches substantially reduced the reported resolution rate.
Unlike these benchmarks, CodeAssay combines taxonomy-first task
construction and hidden-test grading with independent ground-truth auditing and
mutation-based test-suite validation.

%% file: CodeAssay.tex
\section{CodeAssay Design}
\label{sec:construction}

We build CodeAssay taxonomy-first, so that a coverage target derived from
software-engineering knowledge drives task selection and authoring precedes model
evaluation (Figure~\ref{fig:pipeline}).

\subsection{Task Construction}
\label{sec:tasks}

\subsubsection{Task schema and taxonomy.}
\label{sec:schema}
Every task carries a natural-language description, a signature, an author-written
reference solution, and a \texttt{pytest} suite, so a generated solution replaces
the reference without project-level scaffolding. The ten categories of Table~\ref{tab:categories} are informed by the SWEBOK~v4
knowledge areas~\cite{swebok4} and by the domain structure of
BigCodeBench~\cite{bigcodebench2024}. They cover the two quality-related areas
SWEBOK~v4 emphasizes, Software Testing through TCQ and Software Security through
CAT. The categories were defined before the tasks were written, and the uneven
distribution reflects a judgment about each area's breadth
(Section~\ref{sec:threats}). 

A second author who did not write the tasks reviewed the category label of every
task in the structured first-batch record of Section~\ref{sec:audit-protocol} and
confirmed 133 of the 138 (\SI{96.4}{\percent}); the five dissents were discussed
and resolved in favor of the original label.

\begin{table}[!b]
  \centering
  \small
  \caption{The ten task categories ($\Sigma = 185$) and their SWEBOK~v4 areas.}
  
  \label{tab:categories}
  \setlength{\tabcolsep}{5pt}
  \begin{tabular}{@{}llrl@{}}
    \toprule
    \textbf{Abbr.} & \textbf{Category} & \textbf{\#Tasks} &
    \textbf{SWEBOK~v4 area} \\
    \midrule
    ADS & Algorithms \& Data Structures & 30 & Computing Foundations \\
    SPT & String \& Text Processing     & 26 & Software Construction \\
    DPS & Data Processing \& Statistics & 21 & Computing Foundations \\
    FIO & File I/O \& System            & 18 & Software Construction \\
    WNT & Web \& Networking             & 16 & Software Design       \\
    CCA & Concurrency \& Async          & 16 & Software Design       \\
    CAT & Crypto, Auth \& Tokens        & 15 & Software Security     \\
    DBS & Database \& Storage           & 15 & Software Construction \\
    TCQ & Testing \& Code Quality       & 15 & Software Testing      \\
    ODP & OOP \& Design Patterns        & 13 & Software Design       \\
    \midrule
    \multicolumn{2}{@{}l}{\textbf{Total}} & \textbf{185} & \\
    \bottomrule
  \end{tabular}
\end{table}

A candidate is included only if it
falls within one category, its behavior is deterministically assertable through
black-box tests without external services, and it admits a compact
self-contained reference solution. Reference solutions average 20.3 source
lines and a cyclomatic complexity of 4.83, and only one task depends on a
third-party package, which keeps evaluation free of network access.

\subsubsection{Test construction and the public/hidden split.}
\label{sec:tests}
Each task's test functions are partitioned into a \emph{public} subset and a
disjoint \emph{hidden} subset. Only the public tests reach a model, as a
\emph{public-test excerpt} and as repair feedback; no hidden test is shown,
executed, or summarized during generation or repair, so a model cannot fit the
exact tests it is scored on (Section~\ref{sec:repair}).

The partition rule is deterministic and independent of test content. Top-level
\texttt{test\_*} functions are taken in source order and assigned by position,
alternating between the subsets, so earlier and later test functions fall on
both sides. Where the count is odd the extra test goes to the hidden subset,
and the module preamble is copied into both files so each runs standalone. We
verified that the two subsets share no test function in any of the 185 tasks. The
1{,}697 test functions divide into 799 public (median 4) and 898 hidden (median
5), and every task retains at least two on each side. Suites average 9.2 tests
(range 4--20) and
attain \SI{98.5}{\percent} statement and \SI{97.0}{\percent} branch coverage.

\subsection{Audit and Validation}
\label{sec:validation}


\subsubsection{Protocol.}
\label{sec:audit-protocol}
We authored CodeAssay in two batches, 138 candidates first and 50 afterwards,
and a second author who did not write the tasks applied the same audit to all
188. For each task the auditor attacked the reference solution, looked for
behavior the description does not state, and constructed inputs intended to break
the tests, reproducing every counterexample before recording it. Disagreements
between the task author and the auditor were resolved by discussion, escalating
to an author who neither wrote nor audited the task.

Each task was scored on a five-dimension rubric covering reference correctness,
description clarity, category fit, test adequacy, and realism. The structured
per-task record was compiled
while the pool stood at 138 candidates and not extended to the 50 later tasks,
so per-task counts are reported over the 138
rather than over all 188. Three of the 138 were dropped as unfixable, and the
remaining 135 together with the 50 later tasks give the 185 released tasks.
\begin{table}[!b]
\centering
\caption{Defect classes from the structured rubric review of the 138-task pool.
Classes overlap, and 87 of the 138 tasks carry at least one. The weak-tests row
counts suites for which the reviewer demonstrated a passing incorrect solution,
whereas the rubric rated 111 of the 138 revise or reject on test adequacy.}
\label{tab:defects}
\begin{tabular}{@{}lr@{}}
\toprule
\textbf{Defect class} & \textbf{Tasks} \\
\midrule
Reference fails or contradicts its own tests          & 7 \\
Tests require behavior the description does not state & 28 \\
Ill-posed or concurrency-defective task               & 3 \\
Reproduced security flaw in the reference             & 5 \\
Weak (non-discriminative) tests                       & 59 \\
Redundancy with another task                          & 6 \\
\bottomrule
\end{tabular}
\end{table}
\subsubsection{Defects found and resolved.}
\label{sec:audit-findings}
The audit surfaced defects across all five rubric dimensions, several of which
would have distorted scores (Table~\ref{tab:defects}). Every fixable task was
repaired until its reference passes the suite and matches the description, the
description alone suffices to solve the task, and the tests reject the incorrect
solutions the auditor constructed. Strengthening the tests grew the assertion
count from 814 to 1{,}242 over the 135 audited tasks that remain in the
benchmark. With the 50
tasks authored later, the released suites hold 1{,}862 assertions across 1{,}697
test functions, and all 185 references pass.

\subsubsection{What the audit changed.}
\label{sec:audit-effect}
A snapshot of the benchmark taken before the audit contains 135 of the
185 released tasks. We regraded every stored program for those tasks against the
pre-audit and the audited suite, using the full suite in each case and
regenerating no model output. The analysis therefore measures how the audit
changes grading outcomes for a fixed set of programs rather than its effect on
generation or repair, since those programs were produced from the audited
descriptions and public tests. 

Regrading changes 170 of the 1{,}890 correctness labels, or \SI{9.0}{\percent},
and it corrects in both directions: 83 programs pass the pre-audit suite and fail
the audited one, and 87 do the reverse, because repairing descriptions and
references also removed tests that a correct solution could not satisfy.
Aggregate correctness is therefore almost unchanged, $91.0\%$ to $91.2\%$, but
the effect is not uniform across models. Among the standard-prompt models the
measured best-to-worst spread widens from 11.9 to 23.7\%.

\subsubsection{Mutation validation of the test suites.}
\label{sec:mutation}
To check test adequacy independently of the rubric review, we run mutation
analysis over the whole benchmark, reporting a mutation
score rather than coverage because coverage correlates only weakly with
test-suite effectiveness~\cite{inozemtseva2014coverage} whereas mutant detection
correlates with real-fault detection~\cite{just2014mutants}. \texttt{mutmut}
generates mutants
from each reference solution and runs that task's suite against every one.

The 185 references yield 4{,}008 mutants. The full suites detect 3{,}311 of them,
a mutation score of $0.826$, and leave 684 alive across 154 of the 185 tasks,
with a further 13 reported as suspicious and counted as undetected. Because
grading uses only the hidden subset, we repeat the analysis on that subset over
the same mutants. The hidden subset detects 2{,}999 of the 4{,}008 ($0.748$),
retaining \SI{90.6}{\percent} of the detections achieved by the full suites, and
leaves 969 alive across 161 tasks with 40 suspicious. We read both scores as
evidence of fault-detection capability rather than as a demonstration of complete
test adequacy or of a balanced split between the public and hidden subsets. We
did not identify equivalent mutants, so both scores are lower bounds.

\subsubsection{Contamination and leakage audit.}
\label{sec:leakage}
Provenance alone cannot rule out overlap with widely circulated tasks, which
Riddell et al.~\cite{riddell2024contamination} measured between code benchmarks
and training corpora. We computed each task's maximum similarity against all
1{,}138 HumanEval and MBPP tasks,
separately for descriptions and for reference solutions, using a normalized
character-level sequence-matching ratio (Python's \texttt{difflib}), treating any
pair above a prespecified threshold of 0.70 as a candidate overlap. The maximum
observed similarity is \textbf{0.48} for descriptions and \textbf{0.64} for
solutions, so no pair exceeds the threshold. We scope the claim narrowly. The
audit shows \emph{no evidence of derivation} from HumanEval or MBPP. It compares
surface form only, so it does not rule out semantic overlap, overlap with sources
other than these two benchmarks, or presence in any model's proprietary training
data.

\subsection{Scoring and Evaluation}
\label{sec:scoring}
\label{sec:metrics}

The primary metric is functional \emph{correctness} on the hidden tests, a
binary per-task outcome that requires passing \emph{all} of a task's hidden
tests. The quality measures are analyzed only for solutions that pass, and we report
three of them plus one context measure, each in the unit its tool produces. None is rescaled and no
composite is formed, because any weighting would encode a preference we cannot
justify from the data.

\emph{Style} is Style is measured as Flake8 violations per 100 logical lines, reported first across all Flake8 violation codes and then with whitespace-only violation codes excluded.
\emph{Complexity} is the raw mean cyclomatic complexity~\cite{mccabe1976} over a
solution's blocks, with the raw maximum, as Radon produces it.
\emph{Selected static-analysis findings} are Bandit and Semgrep
counts by severity, reported descriptively; we do not combine the tools and form
no security score, and these counts are not a measure of program security
(Section~\ref{sec:threats}). \emph{Length} is the logical lines of code in a
solution, as Radon counts them, reported as context rather than as a
quality measure.

Comparisons of these measures use only tasks solved correctly by every
configuration compared (Section~\ref{sec:jointly}). Missing values are dropped
rather than imputed (Section~\ref{sec:threats}). Mutation adequacy is not a per-solution axis, since
it measures how well a \emph{test suite} kills injected faults.

%% file: ExperimentalSetup.tex
\section{Experimental Setup}
\label{sec:setup}

We evaluate 14 model--prompt configurations on all 185 CodeAssay tasks. The
replication package~\cite{codeassay2026artifact} documents the execution
environment and tool versions, and its released outputs make the reported
results exactly reproducible even though closed models can change behind a
fixed identifier.

\subsection{Research Questions}
\label{sec:rqs}

\begin{description}
  \item[RQ1.] Under the standard prompt, how do the seven models compare in (i) final hidden-test correctness and (ii) the selected code-quality measures, evaluated on the common set of tasks solved correctly by all 14 model–prompt configurations?
  \item[RQ2.] Within each model, does the security-focused prompt change
    functional correctness or the quality measures?
  \item[RQ3.] Under the standard prompt, how does final hidden-test correctness
    vary across task categories?
\end{description}

\subsection{Configurations and Generation Settings}
\label{sec:agents}

\subsubsection{Models and prompts.}
We evaluate seven models from three families. Five ran at temperature~0. GPT-5.6-sol and Claude~Sonnet~5 do not accept that
parameter, so it was omitted and they ran at the provider default, and decoding
is therefore not uniform across the seven (Section~\ref{sec:threats}). All seven
share a 16{,}384-token output limit. We sent no
reasoning, thinking, provider-routing or seed parameter, so those settings also
took provider defaults.

Each model runs under two prompts, which differ only in system message.
Following secure-prompting practice~\cite{secprompt2025}, the
\emph{security-focused} message adds \emph{security-conscious} to the persona
sentence of the \emph{standard} message and appends one instruction,
``Prioritize security: validate all inputs, avoid eval/exec, use the secrets
module for randomness, sanitize external data, and handle errors without exposing
internals.''  Pairing them yields $7\times2 = 14$
\emph{model--prompt configurations}.

Generation issued 2{,}590 initial attempts and 156 repairs. Thirteen runs used a 60-second client timeout. Two attempts under GPT-5.6-sol with the
security-focused prompt reached that limit and returned no initial response, and were reissued at a 900-second timeout, so 2{,}748 attempts returned 2{,}746 responses. Every response carries a completion status or stop reason, and each records that the
model stopped on its own rather than at the output limit, so no response was
truncated and the corpus holds a program for every task and configuration.

\subsubsection{Generation and repair.}
\label{sec:repair}
Generation proceeds in at most two attempts. The model first receives the task
description, the signature, and a \emph{public-test excerpt}, namely the first 80
lines of the public test file. Only 2 of the 185 public files exceed that
length, so 796 of the 799 public test functions appear in full. The attempt is then run against the
complete public suite rather than against the excerpt. If it passes, it is the scored solution. If it fails, the model is called once
more
with the failure report appended to the original prompt, the last 80 lines of
the pytest output cut to its last 1{,}200 characters, and this repaired attempt
replaces the first as the scored solution. There is no
third attempt, no tool use, and no planning step.

The reported outcome is final hidden-test correctness after an initial generation and at most one public-test-guided repair, so it is
not the conventional one-shot \passk{} estimator at $k=1$.
Section~\ref{sec:rq1} reports the two parts separately.
No configuration used more than \SI{3.8}{\percent} of its
10{,}000{,}000-token budget, so no eligible repair was skipped.

\subsection{Statistical Protocol}
\label{sec:stats}

Every test is two-sided. The repair decomposition, the static-analysis
counts over all 185 tasks, and RQ3 are reported descriptively and are not tested.
For RQ3 the category mean is final hidden-test correctness averaged over the
seven standard-prompt models, and the between-model spread within a category is
the difference between the highest and the lowest of those seven.

\subsubsection{Tests for binary per-task correctness.}
Per-task correctness is binary and measured repeatedly on the same 185 tasks, so
signed-rank procedures are not appropriate for it. For RQ1 we test the seven
standard-prompt models for an overall difference with
Cochran's~$Q$~\cite{cochran1950} and compare them pairwise with exact McNemar
tests on the discordant pairs, whose $p$-values are exact
binomial. RQ2 applies the same test within each model between the two prompts.
Each contrast reports the discordant counts and the paired risk difference in
percentage points with a 95\% Wald interval computed from those counts;
per-model correctness carries a Wilson score interval.
Holm--Bonferroni correction is applied within each family
separately, the 21 pairwise contrasts of RQ1 and the seven within-model contrasts
of RQ2, and for each quality measure separately.

\subsubsection{Comparison on jointly solved tasks.}
\label{sec:jointly}
Quality indicators are analyzed only for solutions that pass, so configurations
would otherwise be compared on different task samples, and the extra tasks a
stronger configuration solves may differ systematically from the rest. Every
quality comparison is therefore restricted to tasks solved correctly by both
configurations compared, and it characterizes that jointly solved subset rather
than all generated programs. Table~\ref{tab:main} and the RQ1 pairwise tests use
the set solved by all fourteen, so every cell rests on the same tasks; each RQ2
comparison uses that model's own jointly correct set, which is larger. We compare
paired values with the Wilcoxon signed-rank test~\cite{wilcoxon1945} under the
Pratt method for zero differences, report tied pairs, and give the matched-pairs
rank-biserial correlation as the effect size under the same convention. No
continuity correction is applied. Every compared set holds at least 120 pairs,
above the size at which the implementation stops using the exact null
distribution, so these $p$-values are asymptotic.

%% file: Results.tex
\section{Experimental Results}
\label{sec:results}

\begin{table}[t]
\centering
\small
\caption{Final hidden-test correctness over all 185 tasks, and quality measures
by model--prompt configuration. Quality measures are calculated on the 120 tasks
solved correctly by all fourteen configurations (Section~\ref{sec:jointly}).}
\label{tab:main}
\setlength{\tabcolsep}{4pt}
\begin{tabular}{@{}lrrrrrrr@{}}
\toprule
& \multicolumn{2}{c}{Correctness} & \multicolumn{2}{c}{Style /100 LLOC} & \multicolumn{2}{c}{Complexity} & \\
\cmidrule(lr){2-3}\cmidrule(lr){4-5}\cmidrule(lr){6-7}
Configuration & \% & $N$ & all & excl.\ ws & mean & max & LLOC \\
\midrule
\multicolumn{8}{@{}l}{\emph{Standard prompt}}\\
GPT-4o-mini        & 77.3 & 143 & 20.10 & 4.26 & 4.93 & 5.21 & 16.2 \\
GPT-4o             & 87.0 & 161 & 16.85 & 4.64 & 4.80 & 5.03 & 16.4 \\
GPT-5.6-sol        & \textbf{98.9} & 183 & 0.12 & 0.12 & 5.45 & 5.90 & 20.5 \\
Claude Haiku~4.5   & 96.8 & 179 & 44.50 & 0.57 & 5.50 & 5.79 & 20.5 \\
Claude Sonnet~4.5  & 95.1 & 176 & 42.24 & 0.52 & 5.76 & 6.02 & 20.6 \\
Claude Sonnet~5    & \textbf{98.9} & 183 & 0.23 & 0.23 & 5.19 & 5.47 & 18.0 \\
Gemini~2.5~Flash   & 91.4 & 169 & 5.09 & 3.35 & 5.76 & 6.15 & 21.1 \\
\midrule
\multicolumn{8}{@{}l}{\emph{Security-focused prompt}}\\
GPT-4o-mini        & 75.1 & 139 & 21.66 & 4.84 & 6.60 & 6.88 & 19.8 \\
GPT-4o             & 89.7 & 166 & 18.49 & 5.50 & 5.83 & 6.08 & 18.9 \\
GPT-5.6-sol        & \textbf{98.4} & 182 & 0.27 & 0.27 & 8.08 & 9.05 & 32.3 \\
Claude Haiku~4.5   & 94.6 & 175 & 45.18 & 0.62 & 7.81 & 8.31 & 28.0 \\
Claude Sonnet~4.5  & 95.7 & 177 & 44.81 & 1.40 & 8.43 & 8.78 & 28.4 \\
Claude Sonnet~5    & 97.8 & 181 & 0.33 & 0.33 & 7.41 & 8.72 & 35.9 \\
Gemini~2.5~Flash   & 91.4 & 169 & 5.63 & 4.58 & 9.17 & 9.84 & 31.3 \\
\bottomrule
\end{tabular}
\end{table}

\subsection{RQ1: Correctness and Quality Under the Standard Prompt}
\label{sec:rq1}

\subsubsection{Correctness.}
Correctness under the standard prompt ranges from $77.3\%$ to $98.9\%$
(Table~\ref{tab:main}). Cochran's $Q$ across the seven models rejects equal
per-task correctness ($Q=121.7$, $\mathrm{df}=6$, $p<0.001$), and pairwise exact
McNemar tests with Holm--Bonferroni correction separate 12 of the 21 pairs.

GPT-4o-mini differs from every other model (Holm $p\le0.004$), with paired risk
differences from $-9.7$ to $-21.6$ percentage points, and GPT-4o differs from
four of the five models that score above it (for example $-11.9$~pp against
GPT-5.6-sol, 95\% CI
$[-16.8,-7.0]$) but not from Gemini. GPT-5.6-sol and Claude~Sonnet~5 both
separate from Gemini ($+7.6$~pp each), while no statistically significant
difference was detected between them or against the Claude~4.x pair.
Claude~Sonnet~5 solves 14 tasks Gemini misses and
misses none that Gemini solves, whereas against GPT-5.6-sol it wins 2 and loses
2.

\subsubsection{Repair outcomes.}
Between 0 and 38 tasks failed the public tests on the first attempt and were eligible for the single repair. Comparing the first attempt with the solution, initial hidden-test
correctness spans $74.1\%$ to $98.9\%$ and final correctness $75.1\%$ to
$98.9\%$, so the repair adds between 0 and 3.8 pp, the largest
gain being GPT-4o under the standard prompt ($83.2\%$ to $87.0\%$). It is not
uniformly beneficial: for GPT-4o-mini it converted four
tasks from failing to passing the hidden tests and three in the opposite
direction, so some repairs improved public-test performance while reducing
hidden-test correctness, which is consistent with overfitting to the visible tests.

\subsubsection{Quality among correct solutions.}
\begin{figure}[t]
\centering
\includegraphics[width=\linewidth]{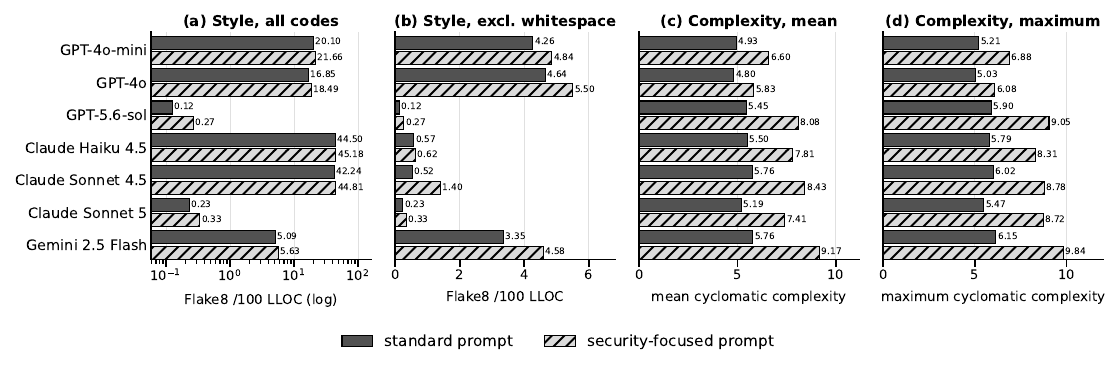}
\caption{Per-model quality measures under the standard (solid) and
security-focused (hatched) prompts, as means over the 120 jointly solved tasks.
Lower is better; style uses a log axis in (a) and excludes whitespace-only
codes in (b).}
\label{fig:metrics}
\end{figure}

Under the standard prompt, no model performs best across all quality measures (Figure~\ref{fig:metrics}). Style separates
the models most, and almost entirely through one kind of violation.
\SI{98.4}{\percent} of Claude~Haiku~4.5's violations and \SI{98.6}{\percent} of
Claude~Sonnet~4.5's are whitespace-only codes, nearly all W293, whitespace on a
blank line, with a small W291 contribution from trailing whitespace on lines that
carry code. Excluding those codes, the two models with the highest total rate
have the third and fourth lowest. The
authored references score $0.12$ violations per 100 logical lines and $4.62$ mean
complexity on the same tasks. Pairwise Wilcoxon tests on the same 120 tasks
separate 19 of the 21 model pairs on style per 100 logical lines, with
matched-pairs rank-biserial correlations from $0.25$ to $1.00$ on the separated
pairs, and 7 of 21 on mean complexity, with correlations from $0.31$ to $0.56$.

Bandit and Semgrep findings were sparse, and their aggregate counts revealed
no consistent ordering. On the 120 jointly solved tasks
each configuration produces 8 to 21 Bandit findings against 12 for the
references, so the models fall on both sides of that baseline, with 8 to 11
medium-severity findings against 10 for the references, 0 to 11 low-severity
findings, and one high-severity finding. Semgrep returned one finding
across the 2{,}590 solutions and 185 references. These findings are descriptive
and should not be interpreted as direct evidence.

In summary, the seven models differ on hidden-test correctness, and 12 of the 21
pairs separate after correction. Among the 120 tasks every configuration solves,
the ordering by correctness is not the ordering by any quality measure, and the
measures differ in how far they discriminate at all, with style separating the
most pairs, mean complexity far fewer, and the static-analysis counts none. The
style result depends further on which violations are counted, since the ranking
inverts once whitespace-only codes are excluded. A style rate therefore describes
a model only together with the violation codes behind it.

\subsection{RQ2: Effect of the Security-Focused Prompt}
\label{sec:rq2}

The security-focused prompt produced no statistically significant change in
correctness for any model. Exact
McNemar tests within each model, corrected across the seven comparisons, leave
every contrast at Holm $p=1.000$. The differences run in both directions and none
exceeds 2.7\%; the largest is a rise, GPT-4o from $87.0\%$ to
$89.7\%$. Gemini~2.5~Flash records $91.4\%$ under both prompts, though the tasks
it solves differ (Section~\ref{sec:threats}). The selected static-analysis
findings move in no consistent direction.
On the 120 jointly solved tasks the Bandit total is identical under both prompts
for three of the seven models and changes for the other four, by up to ten
findings for GPT-4o, and per-task findings change for six of the seven. Over all
185 tasks, low-severity Bandit findings for Gemini~2.5~Flash rise from 0 under
the standard prompt to 20 under the security-focused one, while GPT-4o's total
falls from 24 to 15.

What the prompt changes is the size and shape of the code. The paired comparisons
below use each model's own jointly solved set, larger than the 120-task
set of Table~\ref{tab:main}, so the values differ from that table. Programs get
longer for all models, by between 2.4 and 16.8 logical lines, and raw
cyclomatic complexity rises for all on both the mean and the maximum (Holm
$p<0.001$; matched-pairs rank-biserial 0.78 to 0.99 for length and
0.52 to 0.85 for complexity).

Violations per 100 logical lines move in the worse direction for all seven, but
only Claude~Sonnet~4.5 reaches significance
after Holm correction ($39.41$ to $43.61$, $p=0.019$), and it is also the only
model to separate on the whitespace-excluded rate ($p=0.010$); the remaining six
range from Holm $p=0.079$ to $0.457$. Because programs grow while the rate per line
barely moves, a raw violation count would have recorded a style penalty that is
mostly a length effect.

In summary, the security-focused prompt produced no statistically significant
change in correctness for any model, and no consistent reduction in the selected
static-analysis findings. What it changed reliably is structure: longer programs
and higher raw complexity for all seven models, with style per 100 logical lines
separating one. On this corpus the instruction costs length and branching without
evidence of the benefit it asks for. That does not establish that security did
not improve, which these two analyzers cannot determine.

\subsection{RQ3: Variation Across Task Categories Under the Standard Prompt}
\label{sec:rq3}

Per-category means range from \SI{88.6}{\percent} to \SI{96.6}{\percent} and
between-model spreads from $9.5$ to $40.0$ points (Table~\ref{tab:rq3},
Section~\ref{sec:stats}). No single model reaches \SI{100}{\percent} in every
category.

\begin{table}[t]
  \centering
  \small
  \caption{Per-category correctness under the standard prompt
  (Section~\ref{sec:stats}).}
  \label{tab:rq3}
  \setlength{\tabcolsep}{5pt}
  \begin{tabular}{@{}llrr@{}}
    \toprule
    \textbf{Abbr.} & \textbf{Category} & \textbf{Mean (\%)} & \textbf{Spread} \\
    \midrule
    TCQ & Testing \& Code Quality       & 88.6 & 40.0 \\
    WNT & Web \& Networking             & 91.1 & 37.5 \\
    ODP & OOP \& Design Patterns        & 93.4 & 30.8 \\
    SPT & String \& Text Processing     & 89.0 & 26.9 \\
    DBS & Database \& Storage           & 88.6 & 26.7 \\
    CCA & Concurrency \& Async          & 89.3 & 25.0 \\
    FIO & File I/O \& System            & 92.9 & 22.2 \\
    CAT & Crypto, Auth \& Tokens        & 93.3 & 20.0 \\
    ADS & Algorithms \& Data Structures & 96.2 & 10.0 \\
    DPS & Data Processing \& Statistics & 96.6 &  9.5 \\
    \bottomrule
  \end{tabular}
\end{table}
In summary, correctness varies across the ten categories, and the between-model
spread varies far more than the means do. A model that is competitive on the
benchmark as a whole can therefore be comparatively weak on one category. We
report the observed variation and do not attribute it to an intrinsic property of
the categories, since category size, task complexity and hidden test count are
not controlled here.

%% file: Discussion.tex
\section{Discussion and Implications}
\label{sec:discussion}
The findings show that model comparisons depend on both the property being
measured and the benchmark used to measure it. Under the standard prompt,
functional correctness distinguishes 12 of the 21 model pairs, total Flake8
findings distinguish 19, and mean cyclomatic complexity distinguishes 7.
However, these measures do not produce the same model ordering, and no model
performs best across all of them (Section~\ref{sec:rq1}). The disagreement is not
merely numerical. For example, the total Flake8 rate for the Claude~4.5 models is
dominated by whitespace-only violations, and excluding these codes substantially
changes their relative position. Similarly, the sparse Bandit and Semgrep
findings provide no consistent ordering among the models. An aggregate quality
score must therefore be interpreted according to the specific properties and
rules it measures rather than treated as a general indicator of generated-code
quality.

The security-focused prompt illustrates the same measurement problem. It
produced no statistically significant change in correctness for any model and no
consistent reduction in the selected static-analysis findings. However, it
increased program length and cyclomatic complexity across all seven models
(Section~\ref{sec:rq2}). These results do not establish that the generated
programs became more or less secure because CodeAssay does not include
task-specific security tests, and the selected analyzers detect only particular
classes of findings. Instead, the results show that a general security
instruction can impose measurable structural costs without providing evidence of
the intended security benefit. Security-focused prompting should therefore be
treated as an intervention whose effects require explicit validation rather than
as a safeguard by itself.

The generation-and-repair pipeline further demonstrates why visible and hidden
tests must remain separate. A single public-test-guided repair increased final
hidden-test correctness by between 0 and 3.8 percentage points, but some repaired
programs improved on the public tests while losing hidden-test correctness.
Although this repair analysis is descriptive, the observed pattern is consistent
with possible overfitting to visible feedback. Evaluations should therefore
distinguish initial-generation correctness from final-pipeline correctness,
protect hidden tests from both generation and repair, and examine whether
improvements obtained from visible-test feedback generalize to the hidden suite.

The benchmark-construction findings show that ground truth is itself part of the
measurement instrument. Regrading the fixed set of stored programs against the
pre-audit and audited suites changed 170 of 1{,}890 correctness labels
(\SI{9.0}{\percent}) and changed the best-to-worst model spread from
11.9 to 23.7 percentage points (Section~\ref{sec:audit-effect}). Because the
programs were generated and repaired using the audited task descriptions and
public tests, this comparison measures the audit's effect on grading rather than
its complete effect on generation and repair. Mutation testing provides
complementary evidence that the suites detect injected faults. However, the
surviving mutants show that neither auditing nor mutation analysis establishes
complete test adequacy (Section~\ref{sec:mutation}). Benchmark references, tests,
and expected outputs should therefore be treated as evidence that must be
validated, rather than as unquestionably correct ground truth.

The category-level results further show that aggregate performance can conceal
where model differences occur. Across the 10 task categories, mean correctness
ranges from \SI{88.6}{\percent} to \SI{96.6}{\percent}, whereas the between-model
spread ranges from 9.5 to 40.0 percentage points (Section~\ref{sec:rq3}). A model
that performs competitively on the complete benchmark may therefore remain
comparatively weak for a particular type of task. Evaluations intended to inform
model selection should sample the task domains relevant to the anticipated use
rather than rely only on a benchmark-wide average.

For practitioners, model choice should be based on a
portfolio of measures aligned with the intended context, including functional
correctness, maintainability, security requirements,
and performance on representative task categories.

More broadly, CodeAssay can serve as an evaluation component in AI-augmented software development processes, allowing organizations to establish model-selection criteria and quality gates based on representative tasks rather than relying exclusively on public leaderboard scores.

%% file: ThreatsValidity.tex
\section{Threats to Validity}
\label{sec:threats}

Leakage risk is limited by authoring tasks from a taxonomy and by the audit of Section~\ref{sec:leakage}, which finds no pair above the threshold. A measurement limit persists for
two concurrency tasks (task091, task127), since on stock-GIL
CPython a lock-free counter is observationally identical to a locked one, so black-box tests cannot verify thread-safety. The SWEBOK~v4 mapping is a positioning device rather than a validated classification, and the second-rater category check covered the first batch of 138 candidates but not the 50 tasks authored later (Section~\ref{sec:audit-protocol}). The harness fixes the interpreter hash seed, and all 2{,}590 stored programs were graded twice with identical outcomes. Three stored programs do not parse and fail the correctness gate, so none enters a quality mean.

An earlier version of this study ran five models at a 1{,}024-token output limit while the two reasoning models ran higher, leaving 36 responses unparsable. Regenerating at a uniform 16{,}384-token limit removed the truncation and changed one result. Gemini~2.5~Flash under the security-focused
prompt rose from $83.8\%$ to $91.4\%$, and the apparent correctness cost of that prompt did not persist.

Each quality measure is a single-tool proxy. Static analyzers of this kind detect
known insecure constructions rather than semantic vulnerabilities, so the
near-absence of findings is evidence about the tools and the corpus rather than
about the security of the generated code. Cyclomatic complexity measures
branching rather than maintainability. Two models do not expose a temperature setting, so decoding
is not uniform across all seven.
These newer models may also carry a higher risk of benchmark--training overlap
that a provenance-based audit cannot rule out. Because we score a single solution
with at most one public-test-guided repair, run-to-run variability is not
characterized, and the repair decomposition does not establish why a repair
succeeds or fails. The study covers proprietary models and Python
only, so it does not characterize open-weight systems or higher-temperature
sampling.

%% file: Conclusion.tex
\section{Conclusion}
\label{sec:conclusion}

We presented CodeAssay, a benchmark of 185 tasks authored from a
category taxonomy, which grades generated code on hidden tests, audits and repairs its
own references and tests before any model is scored, and compares quality
measures on the set of tasks every configuration solves correctly. Under the standard prompt the seven models differ significantly on hidden-test
correctness, and no model leads on every quality measure. A security-focused prompt produced no statistically significant change in correctness for any model, while making programs longer and increasing cyclomatic complexity
For practitioners, a green test run and a single leaderboard number understate
the trade-offs in accepting generated code. The protocol is independent of any
single model, so it can be re-instantiated for open-weight models, which we leave to future work.

\subsubsection*{Acknowledgment.}
This work has been supported by FAST, the Finnish Software Engineering Doctoral Research Network, funded by the Ministry of Education and Culture, Finland.

\subsubsection*{Declaration of AI Assistance.} During manuscript preparation, the authors used ChatGPT for grammar, sentence-structure, and formatting refinement; the authors reviewed and revised all content and assume full responsibility.

%% file: references.bib
@article{humaneval2021,
  author  = {Chen, Mark and Tworek, Jerry and Jun, Heewoo and Yuan, Qiming and Pinto, Henrique Ponde De Oliveira and Kaplan, Jared and Edwards, Harri and Burda, Yuri and Joseph, Nicholas and Brockman, Greg and others},
  title   = {Evaluating Large Language Models Trained on Code},
  journal = {arXiv preprint arXiv:2107.03374},
  year    = {2021}
}

@article{mbpp2021,
  author  = {Austin, Jacob and Odena, Augustus and Nye, Maxwell and Bosma, Maarten and Michalewski, Henryk and Dohan, David and Jiang, Ellen and Cai, Carrie and Terry, Michael and Le, Quoc and others},
  title   = {Program Synthesis with Large Language Models},
  journal = {arXiv preprint arXiv:2108.07732},
  year    = {2021}
}

@inproceedings{evalplus2023,
  author    = {Liu, Jiawei and Xia, Chunqiu Steven and Wang, Yuyao and Zhang, Lingming},
  title     = {Is Your Code Generated by {ChatGPT} Really Correct? {Rigorous} Evaluation of Large Language Models for Code Generation},
  booktitle = {Advances in Neural Information Processing Systems (NeurIPS)},
  volume={36},
  pages={21558--21572},
  year      = {2023}
}

@inproceedings{bigcodebench2024,
  author    = {Zhuo, Terry Yue and Vu, Minh Chien and Chim, Jenny and Hu, Han and Yu, Wenhao and Widyasari, Ratnadira and Yusuf, Imam Nur Bani and Zhan, Haolan and He, Junda and Paul, Indraneil and others},
  title     = {{BigCodeBench}: Benchmarking Code Generation with Diverse Function Calls and Complex Instructions},
  booktitle = {Proc.\ International Conference on Learning Representations (ICLR)},  year      = {2025}
}

@inproceedings{livecodebench2024,
  author  = {Jain, Naman and Han, King and Gu, Alex and Li, Wen-Ding and Yan, Fanjia and Zhang, Tianjun and Wang, Sida and Solar-Lezama, Armando and Sen, Koushik and Stoica, Ion},
  title   = {{LiveCodeBench}: Holistic and Contamination Free Evaluation of Large Language Models for Code},
  booktitle={International Conference on Learning Representations},  year={2025}
}

@article{race2024,
  author  = {Zheng, Jiasheng and Cao, Boxi and Ma, Zhengzhao and Pan, Ruotong and Lin, Hongyu and Lu, Yaojie and Han, Xianpei and Sun, Le},
  title   = {Beyond Correctness: Benchmarking Multi-dimensional Code Generation for Large Language Models},
  journal = {arXiv preprint arXiv:2407.11470},
  year    = {2024}
}

@inproceedings{pearce2022,
  author    = {Hammond Pearce and Baleegh Ahmad and Benjamin Tan and
               Brendan Dolan-Gavitt and Ramesh Karri},
  title     = {Asleep at the Keyboard? {Assessing} the Security of {GitHub Copilot}'s
               Code Contributions},
  booktitle = {Proc.\ IEEE Symposium on Security and Privacy (S\&P)},
  pages     = {754--768},
  year      = {2022}
}

@article{sun2026,
  author  = {Sun, Xin and St{\aa}hl, Daniel and Sandahl, Kristian and Kessler, Christoph},
  title   = {Quality Assurance of {LLM}-generated Code: Addressing Non-Functional Quality Characteristics},
  journal = {Journal of Systems and Software},
  volume  = {238},
  pages={112885},
  year    = {2026},
  publisher={Elsevier}
}

@inproceedings{secprompt2025,
  author    = {Bruni, Marc and Gabrielli, Fabio and Ghafari, Mohammad and Kropp, Martin},
  title     = {Benchmarking Prompt Engineering Techniques for Secure Code Generation with {GPT} Models},
  booktitle = {2025 IEEE/ACM 2nd International Conference on AI Foundation Models and Software Engineering (FORGE)},
  pages={93--103},
  year={2025},
  organization={IEEE}
}

@online{mutmut,
  author  = {Anders Hovmöller},
  title   = {mutmut: Mutation Testing Tool},
  year    = {2016},
  urldate = {2026-05-14}
}

@article{mccabe1976,
  author  = {McCabe, Thomas J},
  title   = {A Complexity Measure},
  journal = {IEEE Transactions on Software Engineering},
  volume  = {SE-2},
  number  = {4},
  pages   = {308--320},
  year    = {1976},
  publisher={IEEE}
}

@article{wilcoxon1945,
  author  = {Wilcoxon, Frank},
  title   = {Individual Comparisons by Ranking Methods},
  journal = {Biometrics Bulletin},
  volume  = {1},
  number  = {6},
  pages   = {80--83},
  year    = {1945},
  publisher={JSTOR}
}

@inproceedings{codereval,
  title={{CoderEval}: A Benchmark of Pragmatic Code Generation with Generative Pre-trained Models},
  author={Yu, Hao and Shen, Bo and Ran, Dezhi and Zhang, Jiaxin and Zhang, Qi and Ma, Yuchi and Liang, Guangtai and Li, Ying and Wang, Qianxiang and Xie, Tao},
  booktitle={Proc.\ 46th IEEE/ACM International Conference on Software Engineering (ICSE)},
  year={2024},
}

@inproceedings{classeval,
  title={{ClassEval}: A Manually-Crafted Benchmark for Evaluating {LLMs} on Class-level Code Generation},
  author={Du, Xueying and Liu, Mingwei and Wang, Kaixin and Wang, Hanlin and Liu, Junwei and Chen, Yixuan and Feng, Jiayi and Sha, Chaofeng and Peng, Xin and Lou, Yiling},
  booktitle={Proc.\ 46th IEEE/ACM International Conference on Software Engineering (ICSE)},
  year={2024}
}

@article{hou2024llm4se,
  author  = {Hou, Xinyi and Zhao, Yanjie and Liu, Yue and Yang, Zhou and Wang, Kailong and Li, Li and Luo, Xiapu and Lo, David and Grundy, John and Wang, Haoyu},
  title   = {Large Language Models for Software Engineering: A Systematic Literature Review},
  journal = {ACM Transactions on Software Engineering and Methodology},
  volume  = {33}, number = {8}, pages = {1--79}, year = {2024}, publisher = {ACM},
}

@inproceedings{riddell2024contamination,
  author    = {Riddell, Martin and Ni, Ansong and Cohan, Arman},
  title     = {Quantifying Contamination in Evaluating Code Generation Capabilities of Language Models},
  booktitle = {Proc.\ 62nd Annual Meeting of the Association for Computational Linguistics (ACL)},
  pages     = {14116--14137}, year = {2024}
}

@inproceedings{perry2023insecure,
  author    = {Perry, Neil and Srivastava, Megha and Kumar, Deepak and Boneh, Dan},
  title     = {Do Users Write More Insecure Code with {AI} Assistants?},
  booktitle = {Proc.\ 2023 ACM SIGSAC Conference on Computer and Communications Security (CCS)},
  pages     = {2785--2799}, year = {2023}
}

@online{pytest,
  author  = {Krekel, Holger and Oliveira, Bruno and Pfannschmidt, Ronny and Bruynooghe, Floris and Laugher, Brianna and Bruhin, Florian},
  title   = {pytest: helps you write better programs}, url = {https://docs.pytest.org}, urldate = {2026-07-27}
}

@manual{swebok4,
  title        = {Guide to the Software Engineering Body of Knowledge (SWEBOK Guide), Version 4.0},
  editor       = {Washizaki, Hironori}, organization = {IEEE Computer Society}, year = {2024},
}

@article{cochran1950,
  author  = {Cochran, William G.},
  title   = {The Comparison of Percentages in Matched Samples},
  journal = {Biometrika},
  volume  = {37},
  number  = {3--4},
  pages   = {256--266},
  year    = {1950},
  publisher={Oxford University Press}
}

@inproceedings{siddiq2022securityeval,
  author    = {Siddiq, Mohammed Latif and Santos, Joanna C. S.},
  title     = {{SecurityEval} Dataset: Mining Vulnerability Examples to Evaluate Machine Learning-Based Code Generation Techniques},
  booktitle = {Proc.\ 1st Int.\ Workshop on Mining Software Repositories Applications for Privacy and Security (MSR4P\&S)},
  pages     = {29--33}, year = {2022}, publisher = {ACM},
}

@inproceedings{vero2025baxbench,
  author    = {Vero, Mark and M{\"u}ndler, Niels and Chibotaru, Victor and Raychev, Veselin and Baader, Maximilian and Jovanovi{\'c}, Nikola and He, Jingxuan and Vechev, Martin},
  title     = {{BaxBench}: Can {LLMs} Generate Correct and Secure Backends?},
  booktitle = {Proc.\ 42nd International Conference on Machine Learning (ICML)},
  series    = {PMLR}, volume = {267}, pages = {61344--61390}, year = {2025}
}

@inproceedings{just2014mutants,
  author    = {Just, Ren{\'e} and Jalali, Darioush and Inozemtseva, Laura and Ernst, Michael D. and Holmes, Reid and Fraser, Gordon},
  title     = {Are Mutants a Valid Substitute for Real Faults in Software Testing?},
  booktitle = {Proc.\ 22nd ACM SIGSOFT International Symposium on Foundations of Software Engineering (FSE)},
  pages     = {654--665}, year = {2014}, publisher = {ACM},
}

@inproceedings{inozemtseva2014coverage,
  author    = {Inozemtseva, Laura and Holmes, Reid},
  title     = {Coverage Is Not Strongly Correlated with Test Suite Effectiveness},
  booktitle = {Proc.\ 36th International Conference on Software Engineering (ICSE)},
  pages     = {435--445}, year = {2014}, publisher = {ACM},
}

@inproceedings{xue2026swebenchplus,
  author    = {Xue, Haoran and Aleithan, Reem and Enan, Nafid and Uddin, Gias and Wang, Song},
  title     = {{SWE-Bench+}: Enhanced {LLM} Coding Benchmark},
  booktitle = {Proc.\ 3rd ACM International Conference on AI-Powered Software (AIware)},
  pages     = {332--339}, year = {2026}, publisher = {ACM},
}

@misc{codeassay2026artifact,
  year = {2026},
  url  = {https://github.com/Code-Assay/CodeAssay}
}
